\documentclass[12pt,preprint]{aastex}

\begin{document}
  \title{CO$^+$ in M 82: A Consequence of Irradiation by X-rays}

  \author{M. Spaans} \affil{Kapteyn Astronomical Institute, P.O. Box
  800, 9700 AV Groningen, The Netherlands}\email{spaans@astro.rug.nl}

  \and

  \author{R. Meijerink} \affil{Astronomy Department, University of
  California, Berkeley, CA 94720, United
  States}\email{rowin@astro.berkeley.edu}

  \begin{abstract}
  Based on its strong CO$^+$ emission it is argued that the M 82
  star-burst galaxy is exposed to a combination of FUV and X-ray
  radiation. The latter is likely to be the result of the star-burst
  superwind, which leads to diffuse thermal emission at $\sim 0.7$ keV,
  and a compact hard, 2-10 keV, source (but not an AGN).
  Although a photon-dominated region (FUV) component is clearly present
  in the nucleus of M 82, and capable of forming CO$^+$, only X-ray
  irradiated gas of density $10^3-10^5$ cm$^{-3}$ can reproduce the
  large, $\sim (1-4)\times 10^{13}$ cm$^{-2}$, columns of CO$^+$ that
  are observed toward the proto-typical star-burst M 82. The total X-ray
  luminosity produced by M 82 is weak, $\sim 10^{41}$ erg s$^{-1}$,
  but this is sufficient to drive the formation of CO$^+$.

  \end{abstract}

%
%________________________________________________________________

\section{Introduction}

M 82 is a star-burst galaxy located at 3.9 Mpc (Sakai \& Madore, 1999). It has a
bolometric luminosity of $\sim 4\times 10^{10}$ L$_\odot$ and a large
body of observational data exists for this system. The observed X-ray
luminosity of M 82 is modest at $\sim 4\times 10^{40}$ erg s$^{-1}$
and ROSAT and ASCA data indicate that it has a hard, strongly absorbed
($N_H\sim 10^{22}$ cm$^{-2}$) power-law component and diffuse thermal
contributions at $kT\approx 0.6$ and $0.3$ keV (Moran \& Lehnert 1997).
More recent XMM observations indicate a range of temperatures from 0.2-1.6 keV,
peaking at 0.7 keV, for the diffuse thermal component (Read \& Stevens 2002).
The thermal contributions result from the star-burst superwind
and the star formation itself (supernova heating of the ISM).
There is a compact hard (2-10 keV) X-ray component that is absorbed with a
photon index of 1.7 in the nucleus M 82, and that varies on a 62 days
period (Kaaret, Simet \& Lang 2006a,b).
There is a modest, $4.4\times 10^{39}$ erg
s$^{-1}$, 2-8 keV diffuse component (Strickland \& Heckman 2007). The work of
Moran \& Lehnert (1997) argues that the
intrinsic luminosity of M 82 in the 0.1-10 keV
band is about four times larger when absorption is properly taken into
account, yielding about $10^{41}$ erg s$^{-1}$ in total.
The CO$^+$ chemistry presented here is most sensitive to the softer
diffuse thermal component (which is largely absorbed for our model clouds),
so subtleties in the hard 2-10 keV band do not impact the presented results.
There is no clear evidence of an AGN (i.e., a buried Seyfert nucleus) in M 82.
The X-ray luminosity is about $\sim 0.2$\% of the total
FUV energy budget as measured by the FIR luminosity (Pak et al.\ 2004).

In the center of the late-type galaxy M~82, bright (and optically
thin) CO$^+$ N=2-1, J=5/2-3/2 (236.06 GHz) and J=3/2-1/2 (235.79 GHz)
rotational emission has been observed by Fuente et al.\ (2006).  Based on
combined CN and HCO$^+$ measurements (their Figure 2 and Table 2), the
latter authors argue for a clumpy photon-dominated region (PDR) model
where $\sim 4\times 10^5$ cm$^{-3}$ clumps are embedded in an
inter-clump medium and are exposed to an enhancement in the FUV radiation
radiation field of $G_0\sim 10^4$. This enhancement is with respect to
the average far-ultraviolet (6-13.6 eV) instellar radiation field in the Milky
Way, which enjoys a typical FUV flux of $1.6\times 10^{-3}$ erg s$^{-1}$ cm$^{-2}$.
It is certainly established that
FUV photons produced by the vigorous star formation in M 82 dominate
the state of the molecular gas in its center (e.g., Pak et al.\ 2004
and references therein for ro-vibrational H$_2$ as well as [CII] and
[OI] fine-structure emission), and observations of CO, HCO$^+$ and CN
can be explained as well by a PDR interpretation.  However,
CO$^+$ is under-abundant by at least an order of magnitude in such PDR
models, compared to the observed CO$^+$ columns of
$\sim (1-4)\times 10^{13}$ cm$^{-2}$ (Fuente et al.\ 2006).

In this letter, we investigate whether the large CO$^+$ columns
measured toward M 82 by Fuente et al.\ (2006) can be explained by the
irradiation of molecular gas by the modest X-ray component that M 82
exhibits, without violating the clear merits of PDR physics.

\section{PDR and XDR models}

We have constructed a set of PDR and X-ray Dominated Region (XDR)
models from the codes described by Meijerink \& Spaans (2005),
Meijerink, Spaans \& Israel (2006) and Meijerink, Spaans \& Israel (2007)
in which we vary
both the incident radiation field, density and cosmic-ray ionization rate.
The thermal balance
(with line transfer) is calculated self-consistently with the chemical
balance through iteration. Absorption cross sections for X-rays are
smaller, $\sim 1/E^3$, than for FUV photons. Therefore, PDRs show a
stratified structure while the changes in the chemical and thermal
structure in XDRs are very gradual.  In XDRs, additional reactions
for fast electrons that ionize, excite and heat the gas are
included. The heating efficiency in XDRs is much higher. Since we
focus on galaxy centers, we have assumed the metallicity to be twice
Solar.
We take the abundance of carbon to be equal to that of oxygen,
since the carbon abundance increases faster than oxygen for larger
metallicity. The precise C:O ratio does not affect our general
results, with the exception of O$_2$ and H$_2$O abundances (Bergin et
al.\ 2000, Spaans \& van Dishoeck 2001).
Because the metallicity, even for a system like M 82, is generally poorly
known in the central regions of active star formation, we have also run
models with Solar metallicity. We come back to these in the discussion section.

Our models (Meijerink et al.\ 2007) have $n=10^2-10^{6.5}$~cm$^{-3}$
and $G_0=10^1-10^5$ ($F_X=0.01-160$~erg~s$^{-1}$~cm$^{-2}$).
We adopt a standard size
for our model clouds of 1 or 10 pc. Our X-ray spectrum follows the spectral
characteristics observed for M 82, corrected
for extinction by the observed hydrogen column density ($\sim 10^{22}$ cm$^{-2}$).
This attenuation is important (factor of four) and is the reason why the total
X-ray luminosity is about $10^{41}$ erg s$^{-1}$ (see introduction).
The X-ray spectral shape matters a lot for the penetration of photons in
particular energy bands. Still, for the case of M 82, molecular ion formation
is not very sensitive to the exact spectrum. Overall most of the M 82 X-ray
energy is emitted below a few keV. Now, a purely
soft spectrum would be absorbed quickly in the outer layers of the XDR,
while a purely hard spectrum would penetrate all the way through. But to
first order it is the total energy deposition rate, basically the integral of
the absorbed flux, that drives the chemistry (Meijerink \& Spaans 2005).
Our model clouds, with
$N_{\rm H}\sim 10^{22}$ cm$^{-2}$ per cloud, are chosen such that they absorb the
bulk of the diffuse thermal X-ray flux of M 82, consistent with the significant
absorption that the observations indicate, and thus the dependence on the exact X-ray spectral
shape is modest as far as the produced column densities are concerned.

The considered densities and fluxes are
representative of the conditions in M 82 in terms of PDRs, but only
the models with $F_X\le 10$~erg~s$^{-1}$~cm$^{-2}$ should be typical
of the X-ray background in M 82.
That is, a total X-ray luminosity of $10^{41}$ erg s$^{-1}$ yields
$F_X\approx 0.4$ erg~s$^{-1}$~cm$^{-2}$ for a point source at 50 pc
from a molecular cloud.
We also consider larger values for $F_X$ because the thermal X-ray emission
is diffuse and molecular gas will enjoy a range of distances to ambient
X-ray sources.
For example, Fuente et al.\ (2006) adopt an emission size of $6''$ for
CO$^+$, corresponding to a linear scale of about 100 pc at the distance to
M 82. Imagine then that the interstellar medium of M 82 consists
of a large number of $\sim$ 1-10 pc clouds. If some of these clouds are at
distances of about 1 pc to individual sources of X-ray radiation, then
only about 1\% of the total X-ray luminosity per source would already yield
$F_X\sim 10$ erg~s$^{-1}$~cm$^{-2}$ impinging on these clouds.
At the same time can the covering factor of about a hundred of these clouds
on the sky approach 5-50\%, and thus contribute significantly to the XDR
signal.

\section{Results}

Figure \ref{columns} shows the CO$^+$ column densities for a
$n=10^3-10^4$~cm$^{-3}$ (10 pc cloud) and a $n=10^4-10^{6.5}$~cm$^{-3}$
(1 pc cloud) density range. When comparing the model results with the observations, one
should realize that the observed CO$^+$ column densities are actually
determined for gas densities around $10^5$ cm$^{-3}$ and a rotational
excitation temperature of about 10 K (Fuente et al.\ 2003). We also show
lower density models in figure \ref{columns} because the excitation of
CO$^+$ is quite complicated (Black 1998, St\"auber et al.\ 2006).
We come back to this point in the discussion.
In table 1 an overview is given of all the relevant chemical species and
their ratios.

It is obvious that XDRs allow columns of CO$^+$ that are
comparable to the observed range of $\sim (1-4)\times 10^{13}$
cm$^{-2}$ for modest densities, while $10^5$ cm$^{-3}$ gas requires a
few clouds to be superposed along the line of sight, which seems very
reasonable.  From the abundances shown in figure \ref{PDRandXDR},
where a comparison is made between a typical XDR and PDR model, it is
evident that the CO$^+$ abundance is more than an order of magnitude
enhanced for the same total impinging flux by energy. This is a direct
consequence of the more significant C$^+$+OH$\rightarrow$CO$^+$+H
pathway in XDRs, where large amounts of C$^+$ and OH co-exist to
large depths (Meijerink \& Spaans 2005). Note in this that the
endo-thermic O+H$_2$$\rightarrow$OH+H reaction is driven efficiently at
the high ($\ge 100$ K) gas temperatures that pertain in XDRs even at
large columns (Meijerink \& Spaans 2005), augmented by the vibrational
excitation of H$_2$. The XDR HOC$^+$ abundances are also much larger
than in PDRs and the model column densities are comparable to those of
CO$^+$, consistent with observations. The XDR HCO$^+$/HOC$^+$ column
density ratios are of the order of 20-40 when the total hydrogen
column density is $\le 10^{23}$ cm$^{-2}$, also consistent with
observations (Fuente et al.\ 2006, their table 2).

Similarly, figure \ref{PDRandXDR} and table 1
show that for column densities exceeding $10^{21.5}$ cm$^{-2}$ the CO$^+$/HCO$^+$
column density ratio reaches values of 0.01-0.1 in XDRs, and is boosted relative to PDRs
for the same ambient density and total impinging flux by energy.
Values of 0.01-0.1 can be reached for PDRs as well, but only if the columns are
modest, $\le 10^{21.5}$ cm$^{-2}$. All this compares well with the
4.5-6.5 mag range for individual clumps in the Fuente et al.\ model. Our
adopted cosmic-ray ionization rate is $5\times 10^{-15}$ comparable to
the Fuente et al.\ value. We find (see also Meijerink et al.\ 2006)
that a boost in the formation of CO$^+$ through an
enhanced cosmic ray ionization rate does not occur because direct
ionization of CO is negligible and the C$^+$ abundance is simply too
small beyond the radical region in PDRs to react with OH.

Finally, in their model Fuente et al.\ (2006) require
about 20-40 PDR clumps of $4\times 10^5$ cm$^{-3}$ density and 7 mag
extinction in order to reproduce the observed CN column of
$\sim 10^{16}$ cm$^{-2}$. Figure \ref{PDRandXDR} shows that
XDRs with low impinging X-ray fluxes do not exhibit strongly enhanced
CN abundances (with large $F_X$ they would), but have abundances similar to
or somewhat smaller than PDRs. Table 1 shows
that our PDR model with $G_0=10^{3.5}$ and $n=10^5$ cm$^{-3}$ produces a CN
column of a few $\times 10^{15}$ cm$^{-2}$, and requires several clumps along
the line of sight, consistent with the CO$^+$ requirement at that density.
%We have adopted a density of $10^5$ cm$^{-3}$ for our fiducial model since
%this fits best with the adopted size of 1 pc for the cloud.
The impinging FUV flux of this PDR model is a factor of 10 below the
best fit model of Fuente et al.\ (2005).
The PDR CN/HCN column density ratios of about 4-7 are also consistent with
observations (Fuente et al.\ 2006, their table 2). The XDR CN/HCN column
density ratios are about 80-180.
However, the HCN abundance in an XDR is not strongly boosted at all
for low $F_X$ (see also Lepp \& Dalgarno 1996, their figure 3). Consequently,
the PDR contribution will dominate the observed HCN (as well as CN) signal
and no inconsistency arises.
Our models do not experience the bi-stability effect, where a low and a
high ionization phase co-exist through the interplay of H$_3^+$, S$^+$
and O$_2$ (e.g.\ Boger \&
Sternberg 2006), because gas-grain neutralization is rapid.

\section{Discussion}

The CO$^+$ abundance in X-ray irradiated interstellar gas is boosted,
relative to the FUV irradiation case. The star-burst galaxy M 82 appears to
need only a modest flux of X-rays, consistent with observations, in order to
reproduce the observed CO$^+$ column across its nuclear disk at a density of
$10^3-10^5$ cm$^{-3}$. We conjecture that other star-burst galaxies may
experience similar effects.

The metallicity in the central regions of M 82 is poorly known. Read \& Stevens
(2002) find super-solar abundances for Mg and Si, but near-solar values for
N, O and Fe).
For comparison we have run models with Solar metallicity and the same
density and irradiation conditions. It turns out that Solar metallicity
lowers the abundance of CO$^+$ by about a factor of two to three for
columns less than a few times $10^{22}$ cm$^{-2}$, while the difference is no
more than $\sim 50$\% for columns larger than a few times $10^{22}$ cm$^{-2}$.
This is a direct consequence of the fact that a lower metallicity causes
a lower absorption rate of X-rays and thus a larger total column of material
is needed to build up the same column of ionization driven species like
CO+. Since the bulk of the M 82 X-rays is absorbed in our model, i.e.\
we have total hydrogen columns of more than $3\times 10^{22}$ cm$^{-2}$
(a few clumps), the impact of metallicity variations is modest.
Specifically, for Solar metallcity
and a total hydrogen column density of $2.5\times 10^{22}$ cm$^{-2}$
($\sim 2$ clumps) or $5\times 10^{22}$ cm$^{-2}$ ($\sim 4$ clumps) we find
CO$^+$ columns of $2\times 10^{12}$ cm$^{-2}$ (a factor 2.7 lower) or
$5\times 10^{12}$ cm$^{-2}$ (a factor 1.5 lower), respectively.

\begin{figure}
\unitlength1cm
\begin{minipage}[b]{7.5cm}
\resizebox{7.5cm}{!}{\includegraphics*[angle=0]{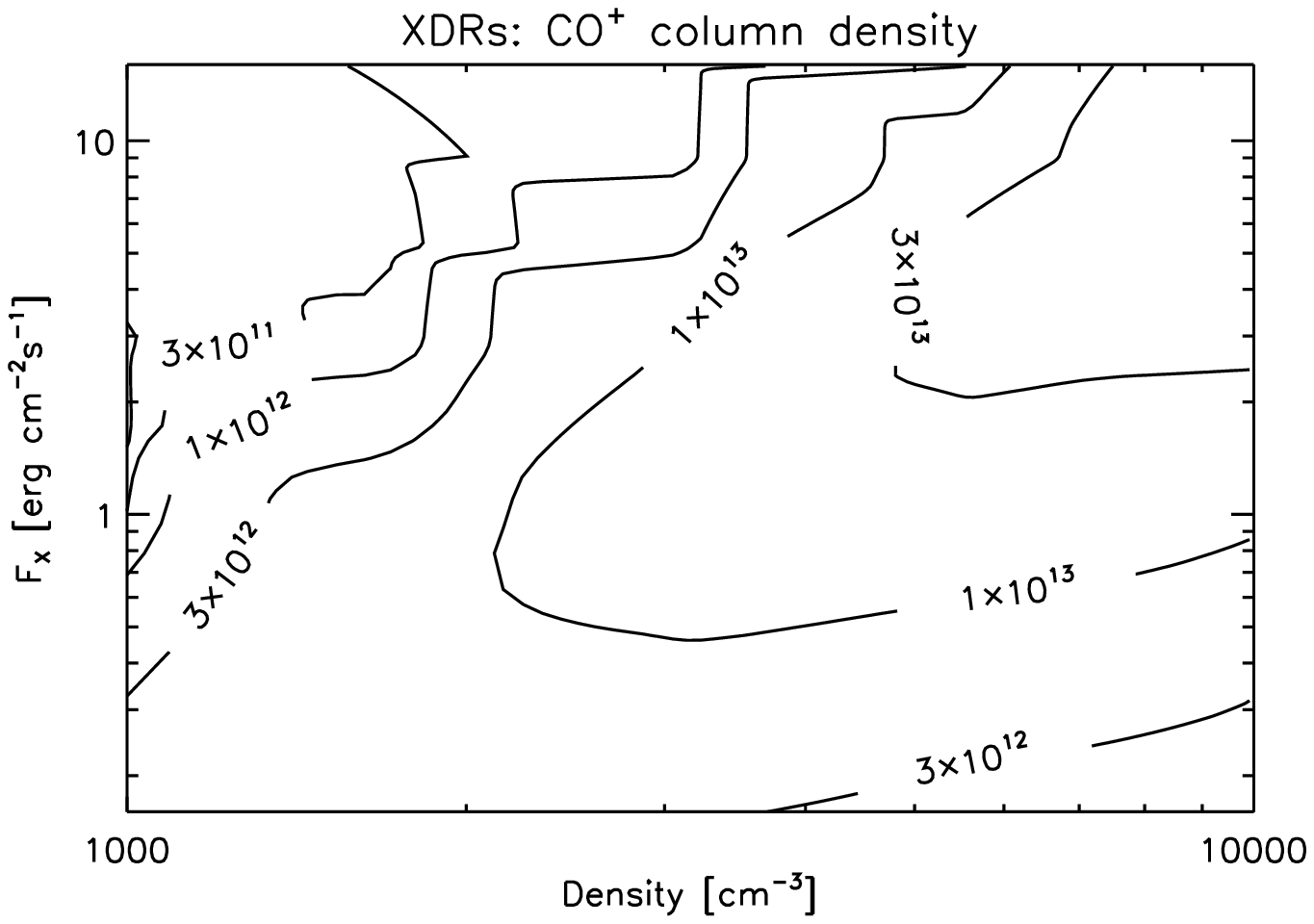}}
\end{minipage}
\begin{minipage}[b]{7.5cm}
\resizebox{7.5cm}{!}{\includegraphics*[angle=0]{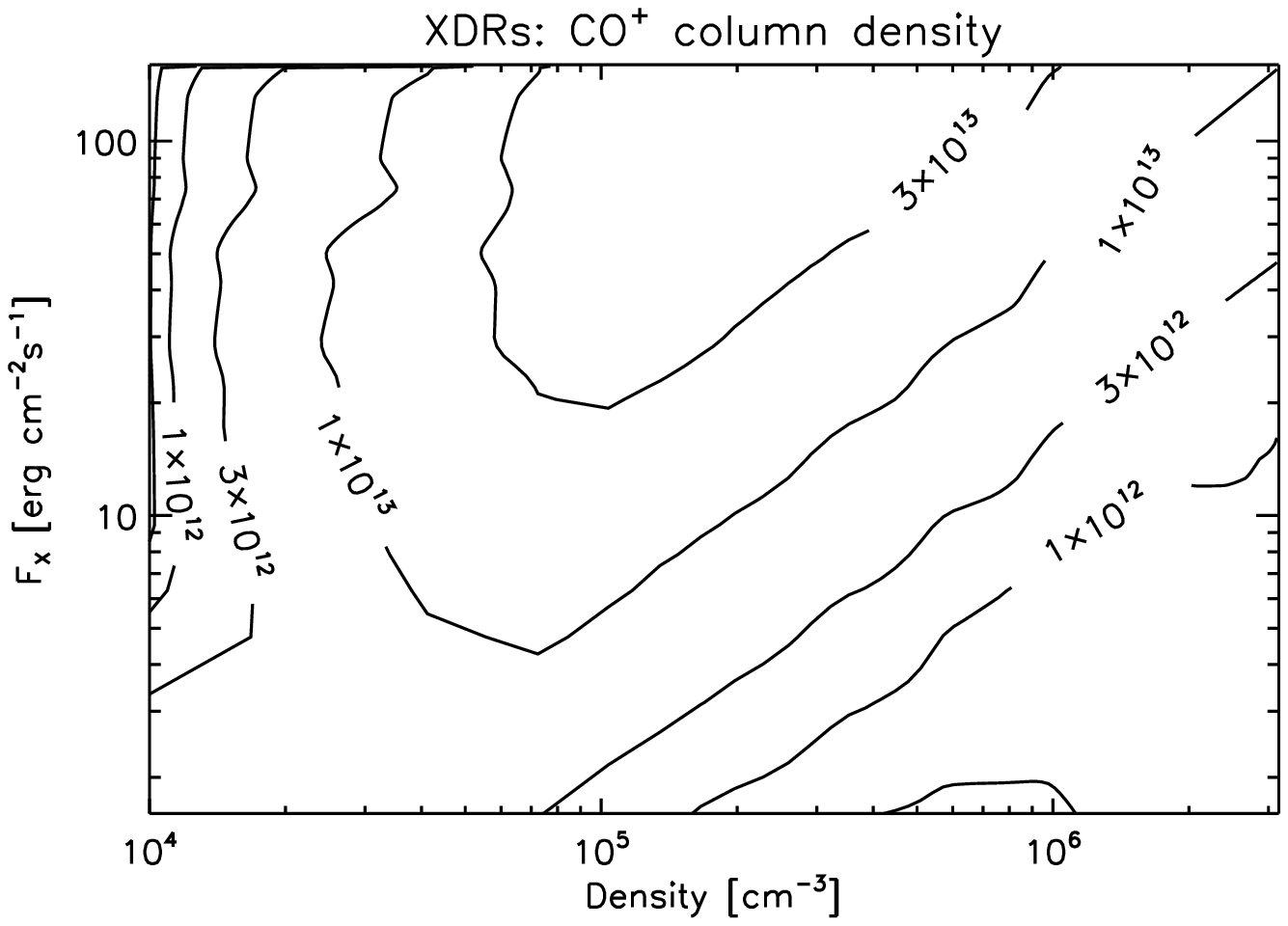}}
\end{minipage}
\caption{CO$^+$ column densities in XDRs as derived from the models
calculated in Meijerink et al.\ 2007. {\it Left:} Mid density range,
$n=10^3-10^4$~cm$^{-3}$, and impinging fluxes
of $F_X=0.16-16$~erg~s$^{-1}$~cm$^{-2}$. {\it Right:} High density range,
$n=10^4-10^{6.5}$~cm$^{-3}$, and impinging fluxes
of $F_X=1.6-160$~erg~s$^{-1}$~cm$^{-2}$.}
\label{columns}
\end{figure}

\begin{figure}
\unitlength1cm
\begin{minipage}[b]{7.5cm}
\resizebox{7.5cm}{!}{\includegraphics*[angle=0]{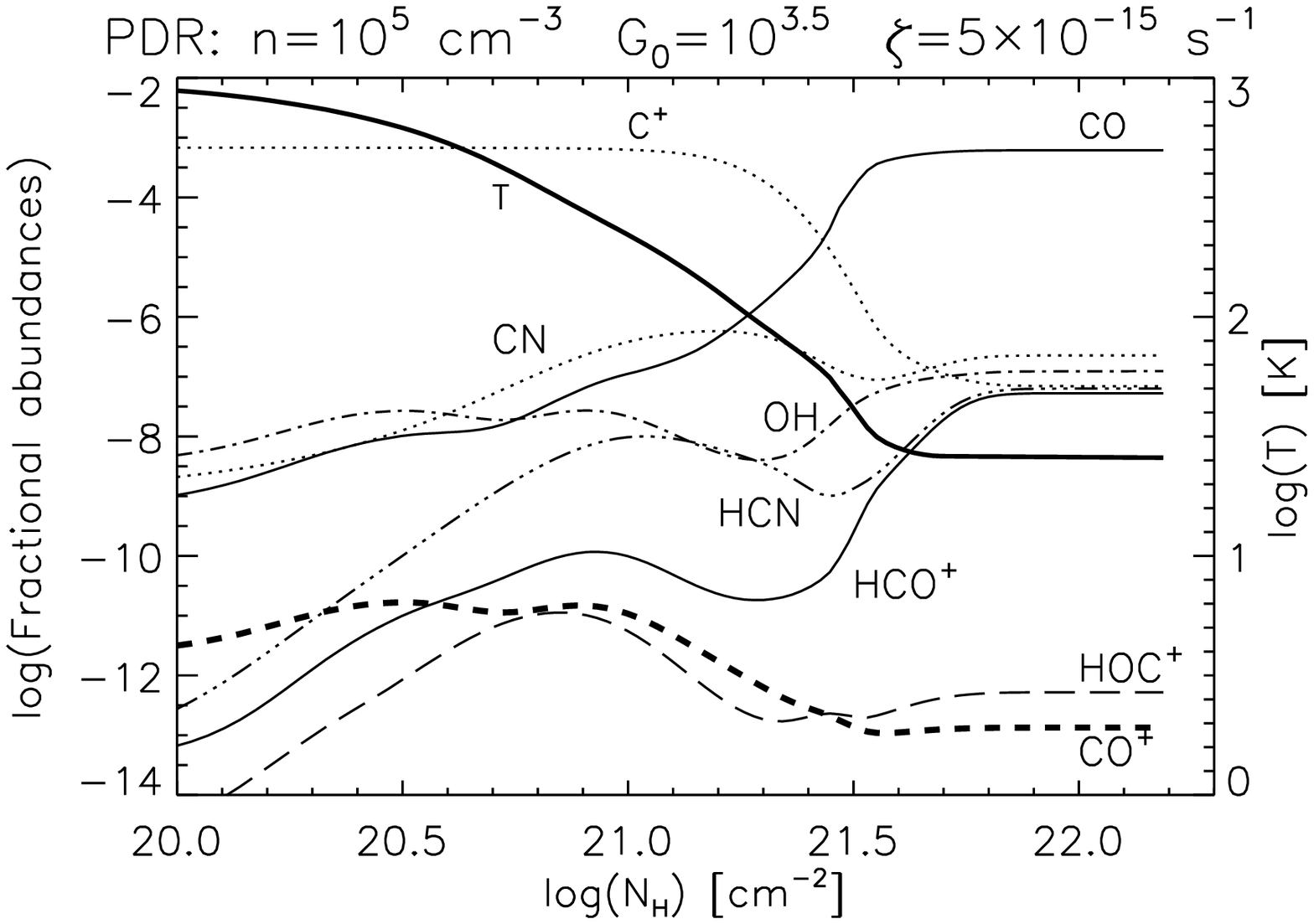}}
\end{minipage}
\begin{minipage}[b]{7.5cm}
\resizebox{7.5cm}{!}{\includegraphics*[angle=0]{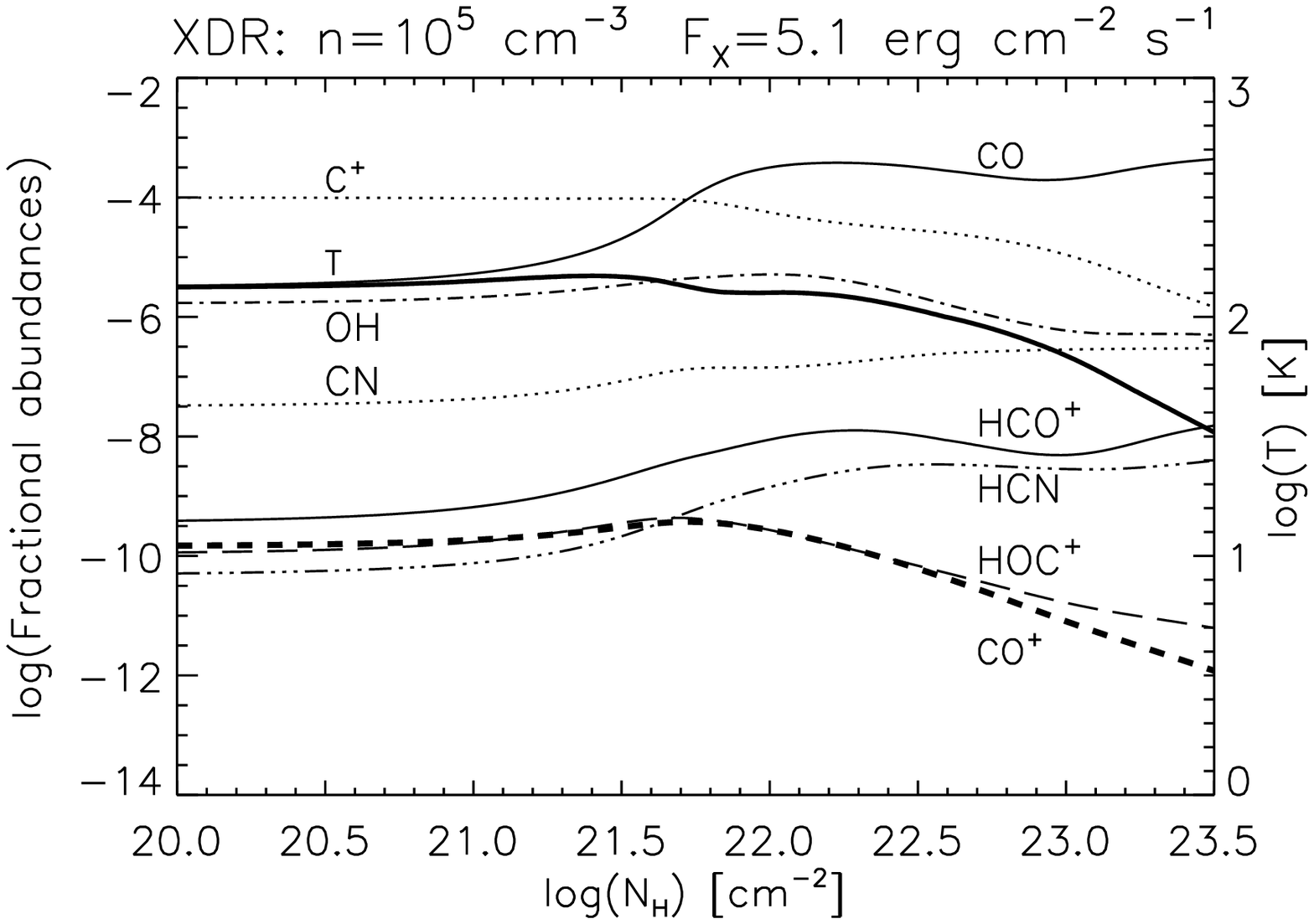}}
\end{minipage}
\caption{Chemical and thermal structure of a PDR with an enhanced
cosmic ionization rate ($\zeta=5\times10^{-15}$~s$^{-1}$) and an XDR
model. Density $n=10^5$~cm$^{-3}$ and
$G_0=10^{3.5}$/$F_X=1.6$~ergs~s$^{-1}$~cm$^{-2}$).
The CO$^+$ abundance is at least an order of magnitude larger in the XDR.}
\label{PDRandXDR}
\end{figure}

It would be quite useful to have H, H$_2$ and electron collisional rate
coefficients for CO$^+$. Still, the situation for CO$^+$ is quite special in
that it is a `transient' molecule (Black 1998), for which the destruction time
is shorter than the time to reach collisional equilibrium. As a consequence,
the excited state that the formation process leaves the CO$^+$ molecule in,
should be an integral part of the excitation analysis because inelastic
collisions with H$_2$ may not be able to thermalize the CO$^+$ levels.
Indeed, if CO$^+$ is formed in an excited state, then a low density,
$n=10^3-10^4$ cm$^{-3}$, gas component as shown in figure
\ref{columns} and table 1 can already
lead to large CO$^+$ emissivities because CO$^+$ column densities increase
with $F_X/n$.

\begin{acknowledgements}
We thank J.P.\ P\'erez Beaupuits, J.\ Mart\'\i n-Pintado and
S.\ Garc\'\i a-Burillo for stimulating discussions on XDR chemistry.
We thank P.\ Kaaret for information on recent X-ray observations of M 82.
\end{acknowledgements}

\begin{table}
\caption{Column densities and column density ratios}
{\footnotesize
\begin{tabular}{|l|l|l|l|l|l|l|l|l|}
\tableline
$N_{\rm H}$  &   N(CO$^+$)  &   N(HOC$^+$)  &  N(HCO$^+$)   &   N(CN)  &  N(HCN)  &  CO$^+$/HCO$^+$  &    HCO$^+$/HOC$^+$   &    CN/HCN \\
\tableline
\multicolumn{9}{|l|}{XDR: $n=10^5$~cm$^{-3}$ and $F_x = 5.1$~erg~s$^{-1}$~cm$^{-2}$} \\
\tableline
1.0e22 &  3.0e12 &  3.3e12 &  4.3e13 &  1.1e15 &  6.0e12 &  0.07 &  13.2 &  181  \\
2.0e22 &  4.8e12 &  5.0e12 &  1.6e14 &  2.7e15 &  2.8e13 &  0.03 &  31.5 &  95.4 \\
3.0e22 &  5.7e12 &  5.9e12 &  2.7e14 &  4.7e15 &  5.9e13 &  0.02 &  46.4 &  78.9 \\
\tableline
\multicolumn{9}{|l|}{XDR: $n=10^{3.5}$~cm$^{-3}$ and $F_x = 1.6$~erg~s$^{-1}$~cm$^{-2}$} \\
\tableline
%1.0e22 &  2.7e10 &  1.8e9  &  3.6e9  &  1.9e12 &  9.0e7  &  7.4  &  2.1  &  2.1e4 \\
%2.0e22 &  3.1e11 &  7.9e10 &  1.7e11 &  1.4e13 &  3.4e9  &  1.8  &  2.1  &  4.2e3 \\
3.0e22 &  1.2e12 &  5.7e11 &  1.5e12 &  5.2e13 &  3.2e10 &  0.8  &  2.6  &  1.6e3 \\
6.0e22 &  8.3e12 &  6.9e12 &  3.7e13 &  5.1e14 &  9.4e11 &  0.2  &  5.4  &  543   \\
9.1e22 &  1.8e13 &  1.5e13 &  1.3e14 &  1.5e15 &  3.8e12 &  0.14 &  8.5  &  400   \\
\tableline
\multicolumn{9}{|l|}{PDR: $n=10^5$~cm$^{-3}$, $G_0 = 10^{3.5}$ and $\zeta=5\times10^{15}$ s$^{-1}$} \\
\tableline
1.0e22 & 1.6e10  & 1.0e10  & 2.8e14  &  2.3e15 & 3.5e14  & 5.6e-5 & 2.8e4 & 6.6 \\
2.0e22 & 1.7e10  & 1.5e10  & 7.8e14  &  4.5e15 & 9.5e14  & 2.2e-5 & 5.2e4 & 4.7 \\
3.0e22 & 1.9e10  & 2.0e10  & 1.3e15  &  6.7e15 & 1.6e15  & 1.4e-5 & 6.5e4 & 4.3 \\
\tableline
%\noalign{\smallskip}
\end{tabular}
}
\label{column_and_ratios}
\end{table}

\clearpage

\clearpage

\end{document}